\pdfoutput=1

\documentclass[11pt, hyperref={hyperfootnotes=true}]{article}

\usepackage{ACL2023}

\usepackage{times}
\usepackage{latexsym}

\usepackage[T1]{fontenc}

\usepackage[utf8]{inputenc}

\usepackage{microtype}

\usepackage{inconsolata}
\usepackage{graphicx,subfigure}
\usepackage{multirow}
\usepackage{makecell}
\usepackage{amssymb}
\usepackage{pifont}
\usepackage{amsmath}
\usepackage{algorithm}
\usepackage{algorithmic}
\usepackage{color}
\usepackage{xcolor}
\usepackage{colortbl}
\usepackage{booktabs}
\usepackage{mathrsfs}
\usepackage{paralist}

\usepackage{xspace}
\newcommand{\ours}{\texttt{NEED}\xspace}

\newcommand{\ie}{\emph{i.e., }}
\newcommand{\eg}{\emph{e.g., }}

\newcommand{\etc}{\emph{etc.}}

\title{{\it Two Heads Are Better Than One:} \\ Improving Fake News Video Detection by Correlating with Neighbors}

\author{Peng Qi$^{1,2}$,  Yuyang Zhao$^{3}$, Yufeng Shen$^{2}$, Wei Ji$^{3}$, Juan Cao$^{1,2}$\thanks{$^*$Corresponding author.}\;
and Tat-Seng Chua$^{3}$
\\
 \textsuperscript{1} Key Laboratory of Intelligent
Information Processing, \\
 Institute of Computing Technology, Chinese Academy of Sciences\\
  \textsuperscript{2} University of Chinese Academy of Sciences\\
  \textsuperscript{3} National University of Singapore\\
  \texttt{\{qipeng,caojuan\}@ict.ac.cn},
  \texttt{yuyang.zhao@u.nus.edu},\\
\texttt{shenyufeng22@mails.ucas.ac.cn},
  \texttt{\{jiwei,dcscts\}@nus.edu.sg}
}

\begin{document}
\maketitle
\begin{abstract}

The prevalence of short video platforms has spawned a lot of fake news videos, which have stronger propagation ability than textual 
fake news. Thus, automatically detecting fake news videos has been an important countermeasure in practice.
Previous works commonly verify each news video individually with multimodal information. Nevertheless, news videos from different perspectives regarding the same event are commonly posted together, which contain complementary or contradictory information and thus can be used to evaluate each other mutually.
To this end, we introduce a new and practical paradigm, 
\ie cross-sample 
fake news video detection, and propose a novel framework, \underline{N}eighbor-\underline{E}nhanced fak\underline{E} news video \underline{D}etection (\ours), which integrates the neighborhood relationship of new videos belonging to the same event.
\ours can be readily combined with existing single-sample detectors and further enhance 
their performances  
with the proposed \textit{graph aggregation} (GA) and \textit{debunking rectification} (DR) modules.
Specifically, given the feature representations obtained from single-sample detectors, GA aggregates the neighborhood information with the dynamic graph to enrich the features of independent samples. After that, DR explicitly leverages the relationship between debunking videos and fake news videos to refute the candidate videos via textual and visual consistency.
Extensive experiments on the public benchmark demonstrate that \ours greatly improves the performance of both single-modal (up to 8.34\%
in accuracy) and multimodal (up to 4.97\% in accuracy) base detectors.
Codes are available in https://github.com/ICTMCG/NEED.
\end{abstract}

\section{Introduction}

\begin{center}
\begin{quote}
\textit{``Listen to both sides and you will be enlightened; heed only one side and you will be benighted.''} \\
\hfill --- Zheng Wei (Tang Dynasty)
\end{quote}
\end{center}

The dissemination of fake news has become an important social issue which poses real-world threats to politics \cite{politics}, finance \cite{finance}, public health \cite{health}, \textit{etc}.
Recently, the prevalence of short video platforms has spawned a lot of fake news videos, which are more convincing and easier to spread compared to textual fake news \cite{bg-fakevideo}.
The Cyberspace Administration of China reported that five of the seven core rumors circulating in the china eastern airlines crash incident originated from short video platforms \cite{donghang}.
Statistics from another study also reveal the powerful propagation of fake news videos, which reports that only 124 TikTok\footnote{\url{tiktok.com}. A popular short-form video sharing platform. } fake news about COVID-19 gained more than 20 million views and 2 million likes, comments and shares,
causing negative influences on millions of people \cite{reporttiktok}. 
%
Therefore, developing automatic detection techniques for fake news videos is urgent to mitigate their negative impact.


\begin{figure}[t]
	\centering
	\includegraphics[width=.95\linewidth]{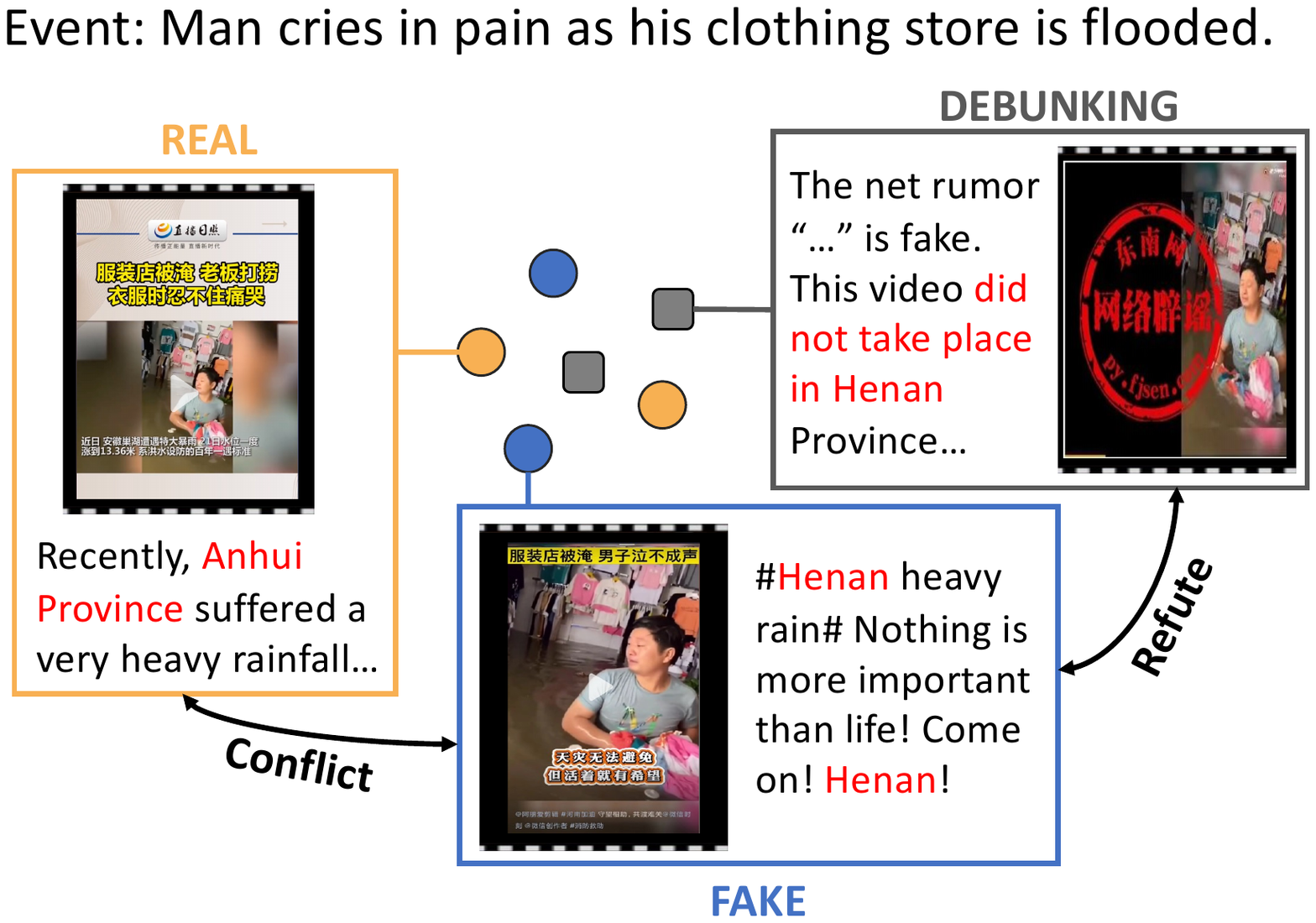}
	\caption{
 A set of videos belonging to the same event. Fake news videos contain conflicting information with the real ones, and the debunking videos can refute the mismatched information in the fake news videos.}
	\label{fig:introcase}
\vspace{-.5em}
\end{figure} 

In view of the practicality of fake news video detection, previous works~\cite{yt-pcancer, yt-covid, myvc, tt, fakesv} leverage the heterogeneous multimodal information of an individual news video for corroboration. 
However, fake news is intentionally created to mislead consumers \cite{surveykdd2017} and thus the multimodal components show few abnormities
after deliberate fabrication. In addition, fake news videos typically contain real news videos where only some frames or the textual description has been maliciously modified to alter their meanings \cite{fakesv}. 
The above characteristics demonstrate that the deliberate fabrication and malicious modification are inconspicuous in a single video, leading to low effectiveness of independent detection by existing works.

In real-world scenarios, when a news event emerges, multiple related videos from different perspectives are posted, including fake news, real news, and debunking videos\footnote{Debunking videos are videos that use factual evidence to refute widely circulated fake news, usually posted by experts.}. Such news videos contain \textit{complementary} or \textit{contradictory} information, which can be used to evaluate each other mutually.
%
As shown in Figure~\ref{fig:introcase}, on the one hand, the fake news video contains conflict information with the real one (\ie different locations: ``Anhui'' province \textit{v.s.} ``Henan'' province). Furthermore, debunking videos also exist in some events, and can easily detect the corresponding fake news by providing fact-based 
\textit{refutations}.
In a newly released dataset \cite{fakesv} based on short video platforms, 54\% of events containing fake news videos also have debunking videos,
but 39\% of events with debunking videos still had fake news videos spread after the debunking videos were posted. 
To some extent, these statistics reveal the universality and insufficient utilization of debunking videos.


Based on the above observations, we conjecture that the relationship among videos of the same event can be modeled to enhance the fake news video
detection and rectify the detection results via factual information.
To this end, we introduce the new cross-sample paradigm for fake news video detection and propose a corresponding novel framework, 
\textbf{\underline{N}}eighbor-\textbf{\underline{E}}nhanced fak\textbf{\underline{E}} news video \textbf{\underline{D}}etection (\textbf{\ours}), which integrates the neighborhood relationship both explicitly and implicitly for better detection.
\ours is a model-agnostic framework, which can easily incorporate various single-sample detectors to yield further improvement. Thus, we first obtain the feature representation from pre-trained single-sample detectors and then refine the representation and final prediction with relationship modeling.

To compensate for the insufficient information in a single video, we organize the news videos in the same event in the form of graph to aggregate the neighborhood information (\textbf{\textit{Graph Aggregation}}). Specifically, we leverage the attention mechanism on the event graph \cite{gat} to model the correlations between different nodes and dynamically aggregate these features.
Furthermore, as mentioned before, there exists explicit relation between debunking and fake news videos, \ie refutations. Consequently, debunking videos can be adopted to rectify the false negative predictions, spotting the ``hidden'' fake news videos (\textbf{\textit{Debunking Rectification}}).
Specifically, we formulate a new inference task to discriminate whether the given debunking video can refute the given candidate video. 
For a given video pair, 
the refutations commonly exist in the textual descriptions of the same visual scenes, which inspires us to detect the textual conflict of the same visual representation. 
To fulfill the discrimination, we take the visual representations from the video copy detector to obtain visual consistency, and fuse it with the textual feature from the textual conflict detector via the attention mechanism. Then the fusion feature is used to classify the refutation relationship between the debunking and candidate videos.
Given the proposed graph aggregation and debunking rectification modules, \ours can significantly improve the performance of base single-sample detectors trained with single-modal or multimodal data. 

Our contributions are summarized as follows:

\begin{compactitem}

\item We propose a new cross-sample paradigm for fake news video detection, modeling multiple news videos in the same event simultaneously. 
Derived from such a paradigm, we propose the \texttt{NEED} framework, which exploits the neighborhood relationship explicitly and implicitly to enhance the fake news video detection.

\item To the best of our knowledge, we are the first to utilize debunking videos in fake news video detection, which can utilize factual information to rectify false negative predictions. To this end, we formulate a new multimodal inference task and propose a novel model that utilizes the consistency from both the textual and visual perspectives to identify whether the given debunking video can refute the given candidate video. 

\item \ours is versatile and can be applied to various single-sample detectors. Extensive experiments on the public benchmark demonstrate that \ours can yield significant improvement with both single-modal and multimodal base detectors.

\end{compactitem}

\section{Related Work}
To defend against fake news, researchers are mainly devoted to two threads of techniques:


\textbf{Fake news detection} methods commonly use non-factual multimodal signals such as linguistic patterns \cite{linguistic}, image quality \cite{icdm, chapter}, multimodal inconsistency \cite{safe, mm21}, user response \cite{defend}, and propagation structure \cite{maprop}, to classify the given news post as real or fake. 
With the prevalence of short video platforms, detecting fake news videos draws more attention in the community.
Recent works mainly leverage deep neural networks to extract the multimodal features and model the cross-modal correlations \cite{myvc, tt, vavd, fakesv}. 
For example, \citet{fakesv} use the cross-attention transformer to fuse news content features of different modalities including text, keyframes, and audio, and use the self-attention transformer to fuse them with social context features including comments and user.   
%
%
%
%

However, existing works in fake news video detection identify each target news independently, without considering the neighborhood relationship in an event.
In view of the practicality of the event-level process,
\citet{naacl} construct a cross-document knowledge graph and employ a heterogeneous graph neural network to detect misinformation.
Nonetheless, this work is performed on the synthetic dataset where each fake news document originates from a manipulated knowledge graph, which cannot be readily applied to real-world scenarios with unpredictable noises in information extraction. Moreover, they only consider the implicit relation among news texts while ignoring the 
explicit refutations between debunking information and fake news.


\textbf{Fact-checking} methods commonly rely on retrieved relevant factual information from reliable sources such as Wikipedia \cite{fever} and webpages \cite{webpage} to judge the veracity of the given check-worthy claim \cite{fcsurvey1,fcsurvey2}. 
A recent thread is to determine whether a claim has been previously fact-checked before retrieving evidence \cite{acl21}. This task is commonly framed as a ranking task, ranking fact-checking articles based on the similarities to the given claim. 
Compared to textual fact-checking, multimodal verification is under-explored. 
\citet{factify} treat the verification as a multimodal entailment task, where the model needs to classify the relationship between the given reliable document (text with associated image) and check-worthy claim (text with associated image).
Inspired by these works, the debunking rectification module in \texttt{NEED} focuses on rectifying the wrong predictions of previously fact-checked news videos by identifying the refutation relationship between the given debunking and candidate news video.


In summary, 
fake new detection methods leverage non-factual patterns learned from large-scale data to give timely judgments for newly emerging events, 
while fact-checking techniques provide more reliable judgments benefiting from the factual information but only work for a part of events limited by the coverage of external sources. 
Our work combines the merits of these two approaches: 
(1) We leverage the data-driven fake news video detectors to obtain effective multimodal representations and to model the neighborhood information, and (2) we also embrace the concept of relevant factual information in fact-checking to rectify the detection results with reliable debunking videos.

\section{Methodology}

\begin{figure*}[t]
	\centering
	\includegraphics[width=\textwidth]{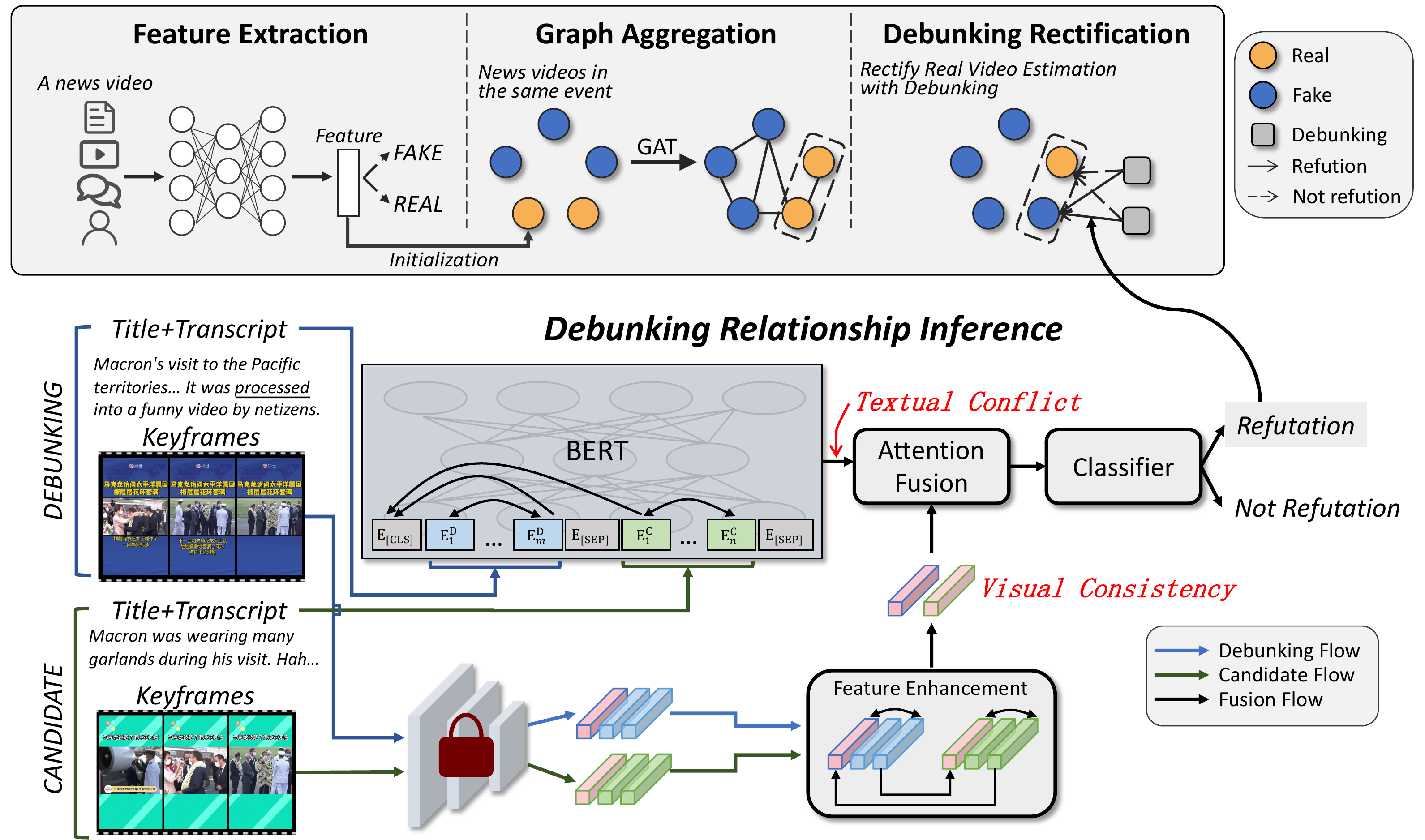}
	\caption{Architecture of the proposed framework \texttt{NEED}. The first row indicates the three stages in \ours, including feature extraction, graph aggregation, and debunking rectification. To realize the debunking rectification, debunking relation inference (the second row) is introduced to determine the refutation relationship.}
	\label{fig:framework}
\vspace{-.1in}
\end{figure*} 

\subsection{Overview}

As mentioned in the Introduction, the fabrication and malicious modification of fake news videos limit the verification ability of existing single-sample fake news video detectors, leading to inferior performance. 
In contrast, the relationship among neighborhood videos, \ie videos of the same event, can be used to supplement the current techniques. Thus, we propose the \underline{N}eighbor-\underline{E}nhanced fak\underline{E} news video \underline{D}etection (\texttt{NEED}) framework, leveraging the set of videos in an event, including fake news $I_{F}$, real news $I_{R}$ and debunking videos $I_{D}$, to improve the performance of single-sample detectors. 
Specifically, \ours is model-agnostic, which takes the representations from the pre-trained base detectors~({Feature Extraction}) to build the dynamic graph and aggregate neighborhood information ({Graph Aggregation, GA}). 
Then, we use the factual information from debunking videos to rectify the predicted results ({Debunking Rectification, DR}). The overall framework is illustrated in Figure~\ref{fig:framework}.


\subsection{Feature Extraction}

News videos contain multimodal information, including title, audio, keyframes, video clips, comments, user profile, \etc \ Existing single-sample fake news video detectors leverage single-modal~\cite{yt-covid} or multimodal~\cite{fakesv} information to discriminate each news video independently. They commonly design tailor-made modules to extract and fuse multimodal features. In contrast, \ours is a solution for the cross-sample paradigm, which can incorporate various single-sample fake news video detectors to yield further improvement with the neighborhood modeling.
Thus, we first extract single-modal/multimodal features $\boldsymbol{F}_{\text{base}}$ for the given set of news videos from the base single-sample detector.

\subsection{Graph Aggregation}
\textbf{Graph Construction.}
Given the set of related news video features $\boldsymbol{F}_{\text{base}}^{E}$ under the same event ${E}$, we organize them in the form of graph attention networks (GAT)~\cite{gat}. 
$\mathcal{G}$ denotes the graph, $\mathcal{V}$ denotes nodes in $\mathcal{G}$ and $\mathcal{E}$ denotes edges between nodes. 
Each node $\boldsymbol{v}_i \in \mathcal{V}$ represents a news video feature from the base detector, and the edge $e_{ij}$ indicates the importance of node $j$’s feature to that of node $i$, which is obtained via attention mechanism.

\noindent\textbf{Feature Aggregation and Classification.}
To aggregate the neighbor information, we apply 
the attention mechanism on the constructed event graph $\mathcal{G}$
to update the representations of nodes. Specifically, given a node $\boldsymbol{v}_i$ with its neighbors $\mathcal{N}_i$, the weight $\alpha_{i,j}$ between $\boldsymbol{v}_i$ and its neighbor $\boldsymbol{v}_j \in \mathcal{N}_i$ is formulated as: 
\begin{equation}
\begin{aligned}
\label{eq1}
    e_{i j} &=\operatorname{LeakyReLU}(\boldsymbol{a}^{\top}[\boldsymbol{W} \boldsymbol{v}_i, \boldsymbol{W} \boldsymbol{v}_j]), \\
    \alpha_{i j} &=\operatorname{softmax}_j(e_{i j})=\frac{\exp(e_{i j})}{\sum_{k \in \mathcal{N}_i} \exp (e_{i k})},
\end{aligned}
\end{equation}
where $\boldsymbol{a}$ and $\boldsymbol{W}$ are trainable parameters, 
$\top$ denotes the matrix transpose,
and $[\cdot,\cdot]$ is the concatenation operation.  
Then, the embedding of $\boldsymbol{v}_i$ is updated by the aggregated information:
\begin{equation}
\hat{\boldsymbol{v}}_i=\sigma(\sum_{j \in \mathcal{N}_i} \alpha_{i j} \boldsymbol{W} \boldsymbol{v}_j),
\end{equation}
where $\sigma$ is the nonlinear operation. 
To avoid over-smoothing of node features, we only adopt two GAT layers. 
The final feature $\hat{\boldsymbol{v}}_i$ is fed into a binary classifier to verify the video. The network is optimized by the binary cross-entropy loss:
\begin{equation}\label{equ:loss}
	\mathcal{L}=-[(1-y)\log (1-p_{\text{GA}}) + y \log p_{\text{GA}}],
\end{equation}
where $p_{\text{GA}}$ is the predicted probability and $y \in \{0,1\}$ denotes the ground-truth label.

\subsection{Debunking Rectification}
Graph aggregation focuses on combining the neighborhood features obtained from base detectors, which learn non-factual patterns from large-scale data.
Instead, there also exists an explicit relationship between fake news videos and debunking videos with factual information, \ie refutations.
Thus, we design the debunking rectification module to rectify the false negative predictions in the previous stages.

Specifically, we propose a new multimodal inference task to recognize this relationship
, \ie debunking relationship inference. The definition of this task is as follows:

\noindent\textbf{Definition 1:} \textit{Given a debunking video and a candidate video that belong to the same event, debunking relationship inference (DRI) aims to determine whether the debunking video can refute the candidate video or not. }

For a given event, we regard videos that are detected to be real by the GA module as the candidates $I_{\rm C}=\{\eta_{\rm C}^1,...,\eta_{\rm C}^{n_{\rm c}}\}$.
For each candidate video $\eta_{\rm C}^i$, we feed it into the DRI model together with the debunking videos $I_{\rm D}=\{\eta_{\rm D}^1,...,\eta_{\rm D}^{n_{\rm d}}\}$ in the same event. 
Then the candidate video is verified by combing the predicted probabilities of graph aggregation $p_{\text{GA}}^i$ and DRI model $p_{\text{DR}}^i$:
\begin{equation}
\begin{aligned}
\label{eq:dr}
    p^i &= \max\{p_{\text{GA}}^i, p_{\text{DR}}^i\}, \\
    p_{\text{DR}}^i &= \mathop{\max}_{\eta_{\rm D}^j \in I_{\rm D}} \texttt{DRI}(\eta_{\rm C}^i, \eta_{\rm D}^j).
\end{aligned}
\end{equation} 



To realize the aim of DRI, we design the model following three principles:
1) Detecting the conflict between the news text of the debunking and candidate videos. 
2) Detecting the consistency between video clips of the given video pair. For example, if the debunking video refutes a piece of fake news that misuses the ``old'' video clip from a previous event, we need to distinguish whether the candidate video uses this ``old'' video clip.
3) Dynamically fusing the textual and visual evidence to eliminate the irrelevant visual information for news events where the visual evidence is not essential, such as ``UN announces Chinese as the international common language''\footnote{As clarified in \citet{unnews}, the fact is that there is no such thing as ``international common language.''}.


Based on the above principles, we propose a novel DRI model, 
which can detect and dynamically fuse textual conflict and visual consistency.

\noindent\textbf{Textual Conflict Detection.} 
Inspired by the task of natural language inference (NLI)~\cite{nli}, we detect the textual conflict via the consistency between the given sentence pair.
Specifically,
given the debunking video, we extract and concatenate the title and video transcript as $S_{\rm D}=[w_1, ...,w_m]$, where $w_i$ represents the $i$-th word in the composed sentence. Likewise, the news text in the candidate news video is represented as $S_{\rm C}=[w_1, ...,w_n]$. 
Then we pack the sentence pair $<S_{\rm D}, S_{\rm C}>$ 
and feed it into BERT 
to model the intra- and inter- sentence correlations.
The BERT we used has been fine-tuned on several NLI datasets to enhance its reasoning ability. 
A learnable type embedding is added to every token indicating  whether it belongs to $S_{\rm D}$ or $S_{\rm C}$. 
Finally, we 
obtain the textual conflict feature:
\begin{equation}
    \boldsymbol{x}_{\rm t} = \texttt{BERT}(\text{[CLS]}S_{\rm D}\text{[SEP]}S_{\rm C}\text{[SEP]}).
\end{equation}

\noindent\textbf{Visual Consistency Evaluation.} 
To match the video clips, we leverage the EfficientNet~\cite{efficientnet} pre-trained on the image similarity dataset~\cite{imgsim} to obtain visual representations of each keyframe. We denote the frame features of the given debunking video and candidate video as $\boldsymbol{F}_{\rm D}=[\boldsymbol{f}^1_{\rm D},...,\boldsymbol{f}^l_{\rm D}]$ and $\boldsymbol{F}_{\rm C}=[\boldsymbol{f}^1_{\rm C},...\boldsymbol{f}^k_{\rm C}]$, respectively.
Following \citet{transvcl}, the fixed sine and cosine temporal positional encoding $\boldsymbol{f}_{\text{tem}}$ are added to the initial features, and a learnable classification token $\boldsymbol{f}^{ \text{[CLS]}}$ is prepended to the feature sequence as the global feature. The processed features of debunking video $\boldsymbol{\hat{F}}_{\rm D}$ and candidate video $\boldsymbol{\hat{F}}_{\rm C}$ are presented as:
\begin{equation}
\begin{aligned}
\boldsymbol{\hat{F}}_{\rm D} &=[\boldsymbol{f}^{\text{[CLS]}}_{\rm D},\boldsymbol{f}^1_{\rm D},...,\boldsymbol{f}^l_{\rm D}]+\boldsymbol{f}_{\text{tem}},
\\
\boldsymbol{\hat{F}}_{\rm C} &=[\boldsymbol{f}^{\text{[CLS]}}_{\rm C},\boldsymbol{f}^1_{\rm C},...,\boldsymbol{f}^k_{\rm C}]+\boldsymbol{f}_{\text{tem}}.
\end{aligned}
\end{equation}

Similar to textual conflict detection, we need to consider intra- and inter- video correlations. Therefore, we employ stacked self- and cross- attention \cite{trm} modules to enhance the initial features, where the query vectors are from the other video in the cross-attention module. 
Finally, the visual consistency feature is obtained by concatenating the classification tokens of the debunking and candidate videos:
\begin{equation}
    \boldsymbol{x}_{\rm v} = [\boldsymbol{f}^{\text{[CLS]}}_{\rm D}, \boldsymbol{f}^{\text{[CLS]}}_{\rm C}].
\end{equation}
 
\noindent\textbf{Attention Fusion and Classification.}
Given the textual conflict feature $\boldsymbol{x}_{\rm t}$ and the visual consistency feature $\boldsymbol{x}_{\rm v}$, we dynamically fuse them to spot the important 
information and eliminate irrelevant information via a self-attention fusion layer. 
Finally, the fused feature is fed into a binary classifier to estimate the probability $p^i_{\text{DR}}$ in Eq.~\ref{eq:dr} that the debunking video can refute the candidate video.

\section{Experiments}
In this section, we conduct experiments to evaluate the effectiveness of \texttt{NEED}. Specifically, we aim to answer the following evaluation questions:
\begin{compactitem}
	\item \textbf{EQ1}: Can \texttt{NEED} improve the performance of fake news video detection?
	\item \textbf{EQ2}: How effective are the different modules of \texttt{NEED} in detecting fake news videos? 
     \item \textbf{EQ3}: How does \texttt{NEED} perform in early detection, which means the number of videos in each event is limited? 
	\item \textbf{EQ4}: How does \texttt{NEED} perform in the temporal split?
\end{compactitem}

\subsection{Experimental Setup}
\noindent\textbf{Dataset.}
We conducted experiments on the FakeSV dataset \cite{fakesv}, the only fake news video dataset that provides rich events and debunking samples.
This dataset collects news videos from popular Chinese short video platforms such as Douyin\footnote{\url{douyin.com}} (the equivalent of TikTok in China), and employs human annotations.
FakeSV consists of 1,827 fake news videos, 1,827 real news videos, and 1,884 debunked videos under 738 events. 
For each news video, this dataset provides the video, title, metadata, comments and user profile. 
Table~\ref{tab:statistics} shows the statistics of this dataset.  

\begin{table}[h]
\small
\centering
\caption{Statistics on the number of news videos in each event.}
\begin{tabular}{lcccc}
\toprule
 & \textbf{\#Fake}   & \textbf{\#Real}     & \textbf{\#Debunking} & \textbf{All}\\
\midrule
Avg. & 3 & 3 & 3 & 8\\
Min. & 0 & 0 & 0 & 1\\
Max. & 24 & 21 & 20 & 25\\
\bottomrule
\end{tabular}
\label{tab:statistics}
\end{table}

\begin{table*}[t]
\small
\centering
\caption{Performance (\%) comparison of base models with and without \texttt{NEED}. The better result in each group using the same base model are in \textbf{boldface}, and the absolute gain is calculated. We report the mean and standard deviation of the five-fold cross-validation.}
\begin{tabular}{l|l|cccccccc}
\toprule
& \multicolumn{1}{l|}{\makecell[c]{\textbf{Method}}} & \multicolumn{2}{c}{\textbf{Acc.}}   & \multicolumn{2}{c}{\textbf{F1}}     & \multicolumn{2}{c}{\textbf{Prec.}}  & \multicolumn{2}{c}{\textbf{Recall}} \\
\midrule
\multirow{8}{*}{\rotatebox{90}{{single-modal}}} & BERT & 77.05$_{\pm3.24}$  & -- & 77.02$_{\pm3.27}$  & -- & 77.21$_{\pm3.12}$ & --& 77.07$_{\pm3.20}$ & --\\
 & \quad + \texttt{NEED} & \textbf{82.99$_{\pm3.86}$} & 5.94$\uparrow$ & \textbf{82.96$_{\pm3.87}$} & 5.94$\uparrow$ & \textbf{83.19$_{\pm3.87}$} &  5.98$\uparrow$ &\textbf{82.99$_{\pm3.88}$} & 5.92$\uparrow$ \\
 
  & Faster R-CNN +Att  & 70.19$_{\pm2.70}$ & -- & 70.00$_{\pm2.68}$ & -- & 70.68$_{\pm2.89}$ & -- & 70.15$_{\pm2.69}$ &-- \\  
 & \quad + \texttt{NEED} & \textbf{78.48$_{\pm3.30}$}& 8.29$\uparrow$& \textbf{78.45$_{\pm3.28}$}& 8.45$\uparrow$& \textbf{78.71$_{\pm3.45}$} & 8.03$\uparrow$& \textbf{78.50$_{\pm3.28}$}& 8.35$\uparrow$\\
  
 & VGGish  & 66.91$_{\pm1.33}$ & -- & 66.82$_{\pm1.30}$& -- & 67.07$_{\pm1.41}$ & -- & 66.89$_{\pm1.32}$ & -- \\
 & \quad + \texttt{NEED} & \textbf{75.25$_{\pm1.61}$} & 8.34$\uparrow$& \textbf{75.12$_{\pm1.63}$} 
 & 8.30$\uparrow$ & \textbf{75.73$_{\pm1.67}$} 
 & 8.66$\uparrow$& \textbf{75.22$_{\pm1.61}$} 
 & 8.33$\uparrow$ \\
  
  & \citet{naacl}     & 
 77.10$_{\pm2.04}$ & -- &
 74.71$_{\pm2.13}$ & -- & 
 76.43$_{\pm2.16}$ & -- & 
 73.98$_{\pm2.05}$ & -- \\
 & \quad + \texttt{NEED} & \textbf{82.96$_{\pm3.42}$}& 5.86$\uparrow$ & \textbf{82.93$_{\pm3.44}$}& 8.22$\uparrow$ & \textbf{83.14$_{\pm3.44}$}& 6.71$\uparrow$ & \textbf{82.95$_{\pm3.46}$}& 8.97$\uparrow$ \\

 \midrule
 
 \multirow{4}{*}{\rotatebox{90}{{Multimodal}}} & FANVM & 76.00$_{\pm2.29}$& -- & 75.98$_{\pm2.30}$& -- & 76.07$_{\pm2.28}$& -- & 76.01$_{\pm2.30}$& -- \\ 
 & \quad + \texttt{NEED} & \textbf{80.97$_{\pm4.05}$}& 
 4.97$\uparrow$& \textbf{80.90$_{\pm4.10}$}& 
 4.92$\uparrow$ & \textbf{81.36$_{\pm3.96}$}& 
 5.29$\uparrow$ & \textbf{80.96$_{\pm4.04}$}& 
 4.95$\uparrow$ \\
  
 & SV-FEND   & 79.95$_{\pm1.97}$ & --& 79.89$_{\pm2.01}$ & --& 80.23$_{\pm1.78}$ & --& 79.94$_{\pm1.98}$& -- \\
 & \quad + \texttt{NEED}  & \textbf{\underline{84.62}$_{\pm2.13}$} & 4.67$\uparrow$ & \textbf{\underline{84.61}$_{\pm2.12}$}
 & 4.72$\uparrow$ & \textbf{\underline{84.81}$_{\pm2.24}$}
 & 4.58$\uparrow$ & \textbf{\underline{84.64}$_{\pm2.14}$}
 & 4.70$\uparrow$ \\

\bottomrule
\end{tabular}
\label{tab:comparison}
\end{table*} 

\noindent\textbf{Evaluation Metrics.}
To mitigate the performance bias caused by the randomness of data split, we follow the setting in \citet{fakesv} and conduct evaluations by doing five-fold cross-validation with accuracy (Acc.), macro precision (Prec.), macro recall (Recall), and macro F1-score (F1) as evaluation metrics. For each fold, the dataset is split at the event level into a training set and a testing set with a sample ratio of 4:1. This ensures that there is no event overlap between different sets, thus avoiding the model detecting fake news videos by memorizing the event information \cite{eann}.

\noindent\textbf{Implementation Details.}
We use two GAT layers in GA and set the hidden states as 128 and 2, 
respectively, with ReLU for the first GAT layer. To avoid overfitting, a dropout layer is added between the two layers with a rate of 0.3.
In DR, we use the pre-trained Erlangshen-MegatronBert-1.3B-NLI\footnote{\url{https://github.com/IDEA-CCNL/Fengshenbang-LM}} to evaluate the textual conflict. For visual consistency evaluation, we use the pre-trained EfficientNet\footnote{\url{https://github.com/lyakaap/ISC21-Descriptor-Track-1st}} to extract the frame features and use the pre-trained weight\footnote{\url{https://github.com/transvcl/TransVCL}} in the feature enhancement module. 
To train the debunking relationship inference model, the debunking videos and fake news videos in the same event are paired with the label ``refutation'', and the debunking videos and real news videos are paired with the label ``not refutation''.
In the attention fusion module, we use a 4-head transformer layer. 
The last two layers of BERT, the visual module and the attention fusion module are trained for 30 epochs with a batch size of 64. The learning rate is set as $1\times10^{-3}$ and $5\times10^{-5}$ in GA and DRI, respectively.
All experiments were conducted on NVIDIA RTX A5000 GPUs with PyTorch.

\subsection{Base Models} \ours can readily incorporate any fake news video detectors that can produce video representation. Here we select four representative single-modal methods and two multimodal methods used in fake news video detection as our base detectors.

\textbf{Single-modal: }
1) \textbf{BERT} \cite{bert} is one of the most popular textual encoders in NLP-related works. We concatenate the video caption and video transcript as a sequence and feed it into BERT for classification.   
2) \textbf{Faster R-CNN+Att}ention \cite{fasterrcnn, trm} is widely used in existing works \cite{tt,fakesv} to extract and fuse the visual features of multiple frames for classification. 
3) \textbf{VGGish} \cite{vggish} is used to extract the acoustic features for classification.
4) \citet{naacl} construct a cross-document textual knowledge graph and employ a heterogeneous graph neural network to detect, which is one of the few works considering the cross-document relationship in fake news detection. 

\textbf{Multimodal:}
1) \textbf{FANVM}~\cite{myvc} use topic distribution differences between the video title and comments as fusion guidance, and concatenate them with keyframe features. An adversarial neural network is used as an auxiliary task to help extract topic-agnostic multimodal features.
2) \textbf{SV-FEND}~\cite{fakesv} use two cross-modal transformers to model the mutual enhancement between text and other modalities (\ie audio and keyframes), and then fuse them with social context features (\ie comments and user) by self-attention mechanism.
Both of these multimodal methods are tailor-made for fake news video detection.

\subsection{Performance Comparison (EQ1)}
We compare the performance of base models with and without \texttt{NEED} in Table~\ref{tab:comparison} and make the following observations:
1) With the help of \texttt{NEED}, all six base models gain significant performance improvement ($4.67\sim8.34$\% in terms of accuracy), which validates the effectiveness and versatility of \texttt{NEED}.
2) 
Compared with \citet{naacl} that combines cross-document information, its basic feature encoder enhanced by \texttt{NEED} (\ie BERT+\texttt{NEED}) achieves better performance, 
verifying the superiority of \texttt{NEED} in utilizing the neighborhood correlations. 
3) 
\ours yields more significant improvement on the underperformed model, \eg 8.34\% improvement in Acc. on VGGish. We conjecture that such a phenomenon can be contributed to the explicit neighborhood modeling in the debunking rectification module, which ensures the lower bound of detection performance via factual information. 

\subsection{Ablation Studies (EQ2)}

\begin{table}[!tbp]
\small
\centering
\caption{Ablation studies on each component in \texttt{NEED}. GA: graph aggregation, DR: debunking rectification. 
The standard deviation values are ignored for simplicity. 
}
\begin{tabular}{lcccc}
\toprule
\multicolumn{1}{l}{\makecell[c]{\textbf{Method}}} & \textbf{Acc.}   & \textbf{F1}     & \textbf{Prec.}  & \textbf{Recall} \\
\midrule
SV-FEND& 79.95 & 79.89 & 80.23 & 79.94\\
\; + DR & 80.94 & 80.90 & 81.15 & 80.93\\
\; + GA & 83.43 & 83.41& 83.61& 83.45\\
\; + \texttt{NEED}~(DR\&GA) & \textbf{84.62} & \textbf{84.61}& \textbf{84.81}& \textbf{84.64}\\
\midrule
VGGish & 66.91 & 66.82 & 67.07 & 66.89\\
\; + DR & 72.84  & 72.70 & 73.30 & 72.84\\
\; + GA & 74.83 & 74.64 & 75.54 & 74.80 \\
\; + \texttt{NEED}~(DR\&GA) & \textbf{75.25} & \textbf{75.12}& \textbf{75.73}& \textbf{75.22}\\
\midrule
DR  &  82.95 & 81.05& 81.36& 81.04\\
\bottomrule
\end{tabular}
\label{tab:ablation}
\end{table}

To verify the effectiveness of each proposed component in \texttt{NEED}, we conduct ablation experiments on top of both SOTA (\ie SV-FEND~\cite{fakesv}) and underperformed (\ie VGGish~\cite{vggish}) models in Table~\ref{tab:comparison}.
%
From Table~\ref{tab:ablation}, we see that
DR and GA consistently improve the performance of both base detectors. 
Moreover, comparing the two enhanced models, DR is more effective on the underperformed model than the SOTA model, 
which supports the explanation that DR ensures the lower bound of detection performance. 

Interestingly, the improvement of DR is less significant than GA, especially on the SOTA model. We conjecture the reason lying in the limited debunking videos, which are only available in 51\% of events in the FakeSV dataset. 
To further verify the effectiveness of factual information introduced by DR, we experiment with DR on the subset that contains debunking videos. Specifically, $p_{\text{DR}}^i$ in Eq.~\ref{eq:dr} is used as the probability that the candidate video is fake.  
As shown in the last row in Table~\ref{tab:ablation}, solely using DR can achieve an accuracy of 82.95\% on the subset, verifying the strong discriminability of debunking videos in detecting fake news videos. 
All the above results demonstrate that the neighborhood relationship can enhance and rectify fake news video detection.



\begin{figure}[!tbp]
    \centering
    \includegraphics[width=.9\linewidth]{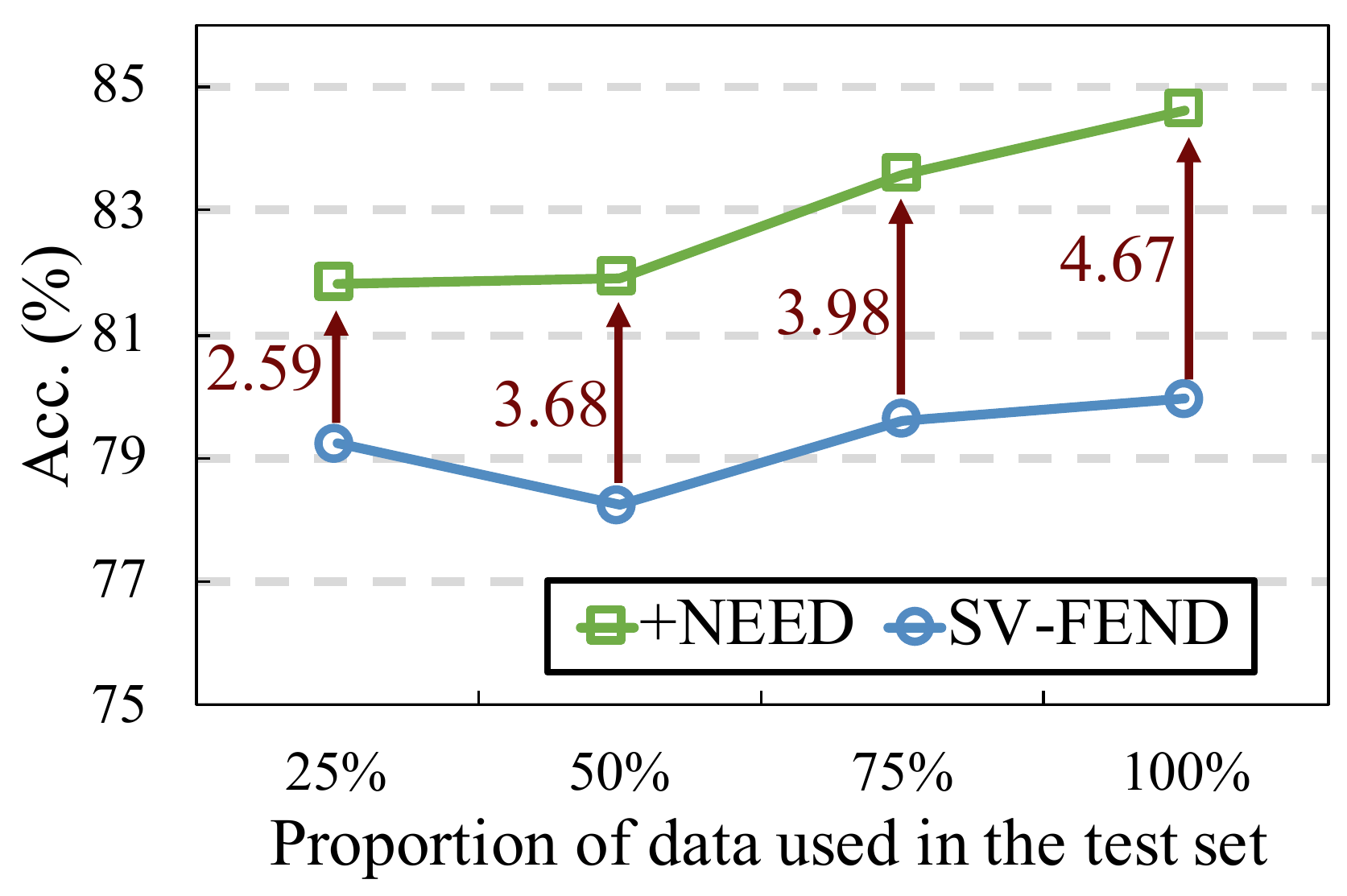}
    \caption{Performance of \texttt{NEED} in early detection. }
    \label{fig:early}
\end{figure}

\begin{table}[!tbp]
\setlength{\belowcaptionskip}{.1em}
\small
\centering
\caption{Peformance of \texttt{NEED} under the temporal split.
} 
\begin{tabular}{lccccc}
\toprule
\makecell[c]{\textbf{Method}} &  \textbf{Acc.}   & \textbf{F1}     & \textbf{Prec.}  & \textbf{Recall} \\
\midrule
\multirow{1}{*}{SV-FEND} & 82.20 & 81.47 & 82.89 & 80.99 \\
\multirow{1}{*}{\quad+\texttt{NEED}}   & {\bf 89.67}  & {\bf 89.37}  & {\bf 90.16}  & {\bf 88.97} \\
\bottomrule
\end{tabular}
\label{tab:temporal}
\end{table} 

\vspace{.3in}
\subsection{Practical Settings}
\noindent\textbf{Early Detection (EQ3).}
Detecting fake news in its early stage is important for timely mitigating its negative influences \cite{survey}. In this part, we conduct experiments using different data proportions of the test set to evaluate the performance of \texttt{NEED} with limited neighbors.
Specifically, we keep the first 25\%, 50\%, 75\% and 100\% videos in each test event in chronological order, and conduct experiments on top of the SOTA base model SV-FEND. Figure~\ref{fig:early} shows that \texttt{NEED} improves the base model even though with limited neighbors. 
Furthermore, as the number of videos within an event increases, \texttt{NEED} yields more significant improvement (from 2.59\% at 25\% data to 4.67\% at 100\% data), 
benefiting from the richer neighborhood relationship.

\noindent\textbf{Performance in Temporal Split (EQ4).}
Splitting data at the event level helps models learn event-invariant features and thus benefit generalization on new events, which is a common practice in the community \cite{eann, mm21}. But in real-world scenarios, when a check-worthy news video emerges, we only have the previously-emerging data to train the detector. Thus we provide another temporal data split, which means splitting the dataset into training, validation and testing sets with a ratio of 70\%:15\%:15\% in chronological order, to evaluate the ability of models to detect future fake news videos.
%
%
Table~\ref{tab:temporal} shows the performance of SV-FEND with and without \ours in the temporal split.
We can see that \ours significantly improves the base model by 7.47\% in Acc., demonstrating that the neighborhood relationship learned by \ours can readily benefit the detection of future fake news videos.


\subsection{Case Studies}

\begin{figure}[!tbp]
	\centering
	\includegraphics[width=.9\linewidth]{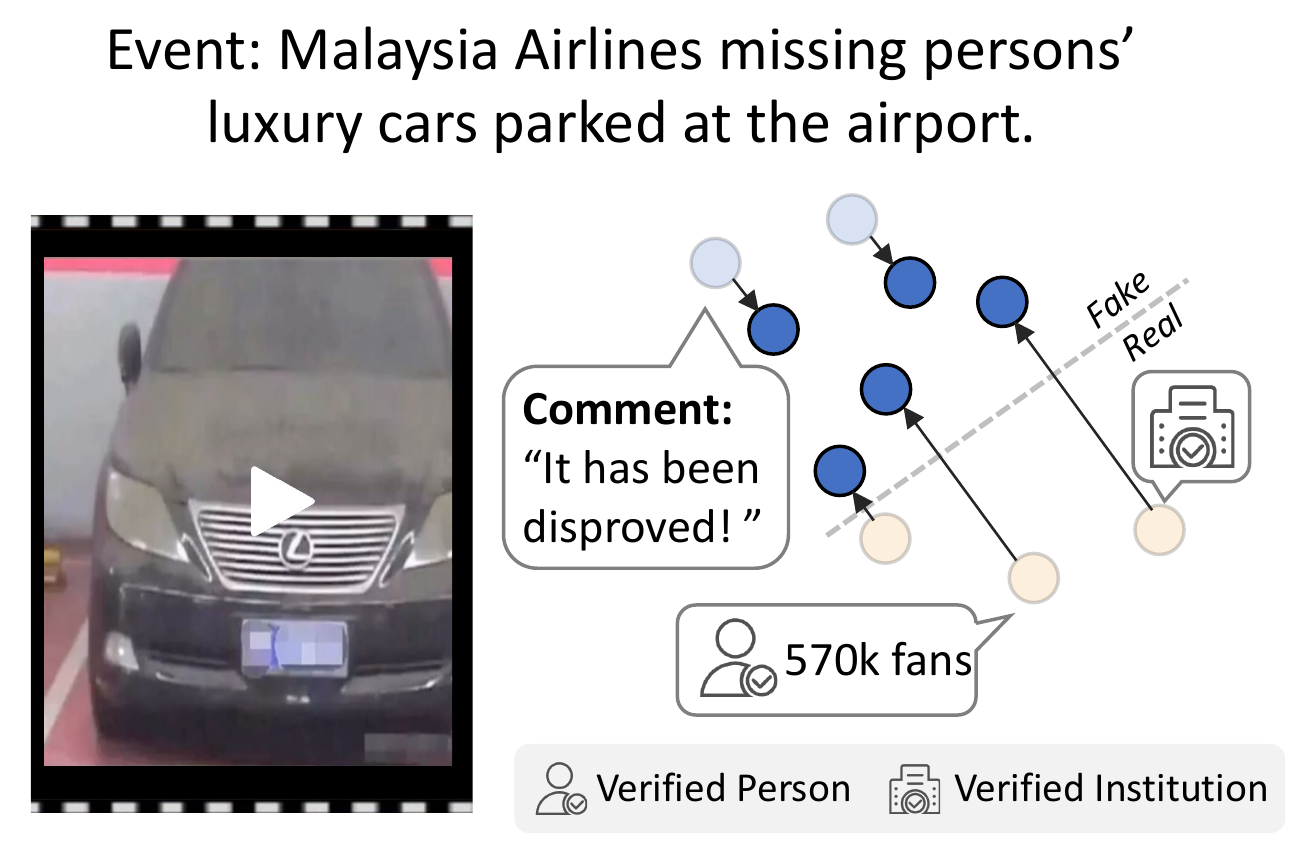}
	\caption{Illustration of the effect of graph aggregation. The left one is the video used in this event, and the right graph shows the score transformation of fake news videos before and after using GA.} 
	\label{fig:case-gnn}
\end{figure}

In this part, we list some cases to intuitively illustrate the effect of GA and DR. 

\noindent\textbf{Graph Aggregation Compensates Single Video Information.}
A single news video contains limited information, and the representation from single-sample detectors can be biased to some data patterns, such as verified publishers.
Figure~\ref{fig:case-gnn} shows the score transformation of multiple fake news videos in the same event before and after using GA. We infer that GA helps by transferring the key clue, \ie the indicative comment, in a single video to others. Moreover, by combining the neighbor information, GA mitigates the publisher bias of single-sample detectors (\ie videos published by verified users are commonly considered to be real). 

\begin{figure}[!tbp]
	\centering
	\includegraphics[width=\linewidth]{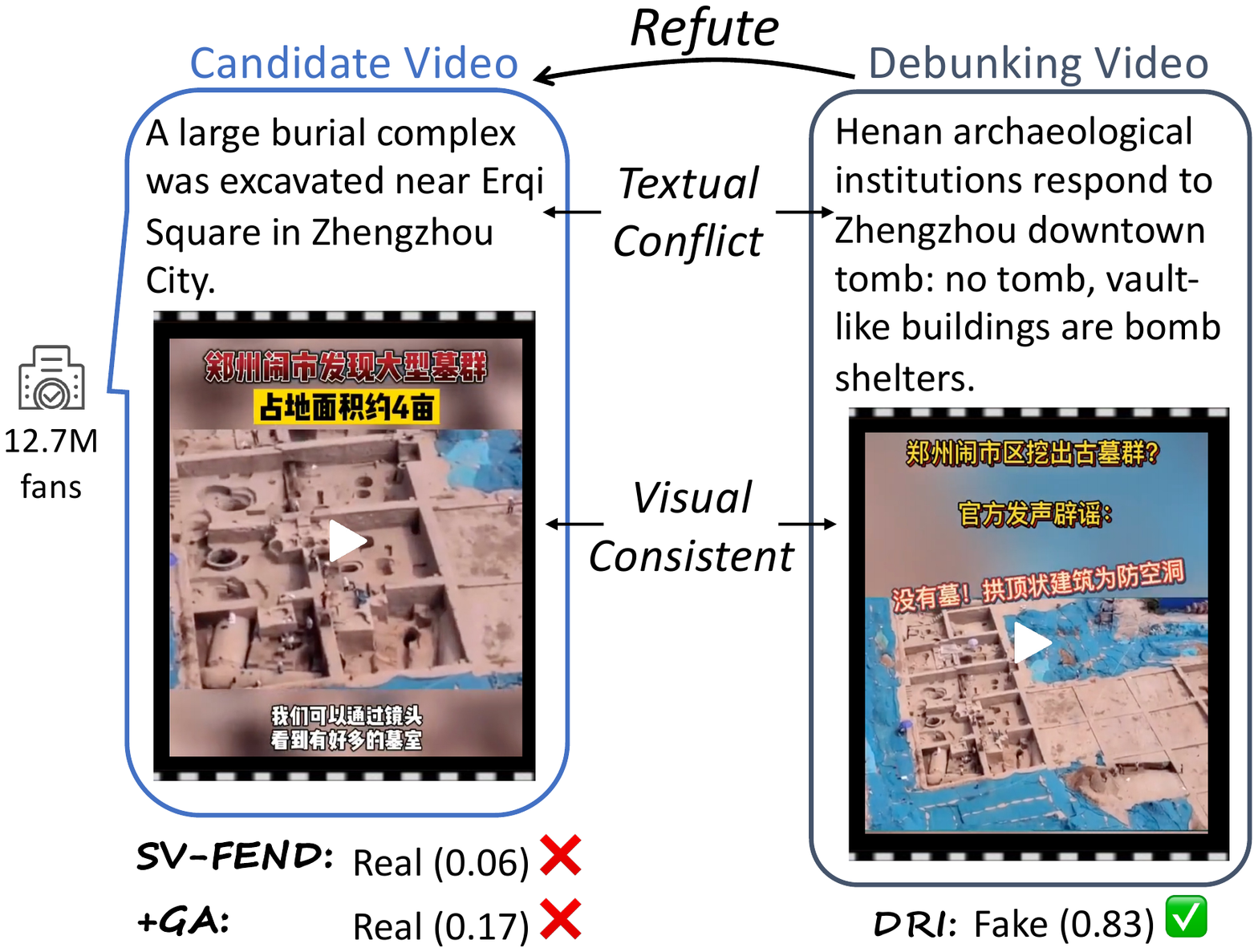}
	\caption{An example where the debunking video helps spot the ``hard'' fake news video missed by previous modules. The number denotes the predicted probability of labels being 0 (real) and 1 (fake), respectively. } 
	\label{fig:case-debunk}
\vspace{-.15in}
\end{figure}

\noindent\textbf{Debunking Rectification Refutes Candidates via Factual Evidence.}
As shown in Figure~\ref{fig:case-debunk}, despite aggregating neighbor information ameliorates the biased prediction (probability 0.06 $\rightarrow$ 0.17) based on the powerful publisher (verified institutional account with 12.7M fans),
GA fails to address such a hard case with a strong bias. Instead, DR uses the debunking video with factual evidence to refute the candidate video, which successfully rectifies the false negative prediction.  

\section{Conclusion}
We proposed a novel framework, namely \texttt{NEED}, to utilize the neighborhood relationship in the same event for fake news video detection. We designed the graph aggregation and debunking rectification modules to assist existing single-sample fake news video detectors. 
Experiments show the effectiveness of \texttt{NEED} in boosting the performance of existing models. We also drew insights on how the graph aggregation and debunking rectification contribute to fake news video detection.



\section*{Limitations}
This work requires that news videos are organized into different events and each event has more than one candidate video. The debunking rectification module  relies on the existence of labeled debunking videos, and the graph aggregation module relies on existing fake news detectors to provide the initial features for each video. The textual length in videos is limited due to that the debunking inference module is based on a pre-trained BERT model with limited sequence length.

%

\section*{Ethics Statement}
Our framework in general does not create direct societal consequences and is intended to be used to defend against fake news videos. 
It can be easily combined into fake news video detection systems, especially when the events have multiple related news videos and debunking videos. 
To the best of our knowledge, no code of ethics was violated throughout the experiments done in this article. Experiments are conducted on the publicly available dataset and have no issues with user privacy.

%

\section*{Acknowledgements}
This work was supported by the National Natural Science Foundation of China (62203425), the Zhejiang Provincial Key Research and Development Program of China (No.2021C01164), the Project of Chinese Academy of Sciences (E141020), the Innovation Funding from the Institute of Computing Technology, the Chinese Academy of Sciences under (E161020).

\bibliography{anthology}
\bibliographystyle{acl_natbib}

\appendix



\end{document}